****** 16.02.2020 26.02.2020 **********

\documentclass[12pt]{article}
\usepackage{blois,graphicx}

\newcommand{\href}[2]{ \, #2}

\begin{document}

\title{Baryon Production at LHC Experiments: Average $p_t$ of Hyperons vs. Energy}

\author{Olga I. Piskounova}

\address{P.N.Lebedev Physics Institute of Russian Academy of Science, Leninski prosp. 53, 119991 Moscow, Russia}

\maketitle\abstract{ 
This paper examines the transverse momentum spectra of baryons in the multi particle production at modern colliders in the frameworks of Quark-Gluon String Model (QGSM). 
It discusses: 1) the difference in $\Lambda^0$ hyperon spectra at proton-antiproton vs. proton-proton reactions on previous colliders; 2) the difference in hyperon spectra 
between the experiments on colliders of low energies and the results from modern machines; 3) the growth of average transverse momenta of $\Lambda$ hyperon with the 
energies of proton-proton collisions up to $\sqrt{s}$ = 7 TeV of LHC experiments. This analysis of baryon spectra led to the following conclusions. First, the fragmentation of 
antidiquark-diquark side of one-pomeron diagram makes the major contribution to baryon production spectra in the asymmetric $p$-$\bar{p}$ reaction. 
Second, the average $p_t$'s of hyperons in p-p collisions steadily grow with energy on the range from $\sqrt{s}$= 53 GeV to 7 TeV. The additional conclusion is the following: since no dramatic changes 
have been seen in the characteristics of baryon production, the hadroproduction processes do not cause the "knee" in the cosmic ray proton spectra at the energies between Tevatron collider and LHC.

Keywords: LHC experiments, spectra of  baryon production, hyperon, cosmic ray proton spectra, average transverse momentum, Quark-Gluon String Model.
}

\section{Introduction}

The aim of this paper is to analyze the transverse momentum spectra of hadrons from the modern collider experiments (ISR \cite{isr},STAR \cite{star},UA5 \cite{ua5}, UA1 \cite{ua1}, CDF \cite{cdf}, ALICE \cite{alice}, ATLAS \cite{atlas} and CMS \cite{cms}). Four reasons warrants this study. 
First of all, The similar analysis of spectra at the previous generation of colliders with lower energies has show the universal form of spectra that is independent of collision energy. Here we have to check wether or not the transverse energy spectra are of similar slope in the range of higher energies.  
Secondly, the preliminary compilation of data on $\Lambda^0$ hyperon transverse momentum distributions \cite{star,ua5} have  demonstrated a difference in the dynamics of multi particle production in proton-proton vs. antiproton-proton collisions. We  suggested that the baryon spectra are sensitive to the asymmetrical reactions. Will this asymmetry work for the spectra at the energies of modern colliders?
Thirdly, a detailed study of characteristics of baryon spectra is necessary at the energies between Tevatron and LHC, because the cosmic ray proton spectrum has a "knee" in this range of energies \cite{knee}. The change in the slope of spectrum of protons, which are produced in the collisions of cosmic rays with the matter of atmosphere, either have an astrophysical origin or can be explained by the substantial change in the dynamics of hadron production at the energies of LHC. 
Finally, the prediction for the behavior of transverse momentum spectra  is needed to plan the future colliders of very high energies.

The Quark-Gluon String Model [QGSM] approach is applied here to the description of $p_t$ spectra for all available flavors of baryons \cite{qgsmtheory}. The Model has successfully described the large volume of data from previous generation of colliders up to the energies $\sqrt{s}$= 53 GeV in the area of low $p_t$ values \cite {veselov}. Recently, $\Lambda^0$ hyperon production have been studied  in updated version of QGSM \cite{lambdaknee}.  

\begin{figure}[htpb]
  \centering
  \includegraphics[width=12.0cm, angle=0]{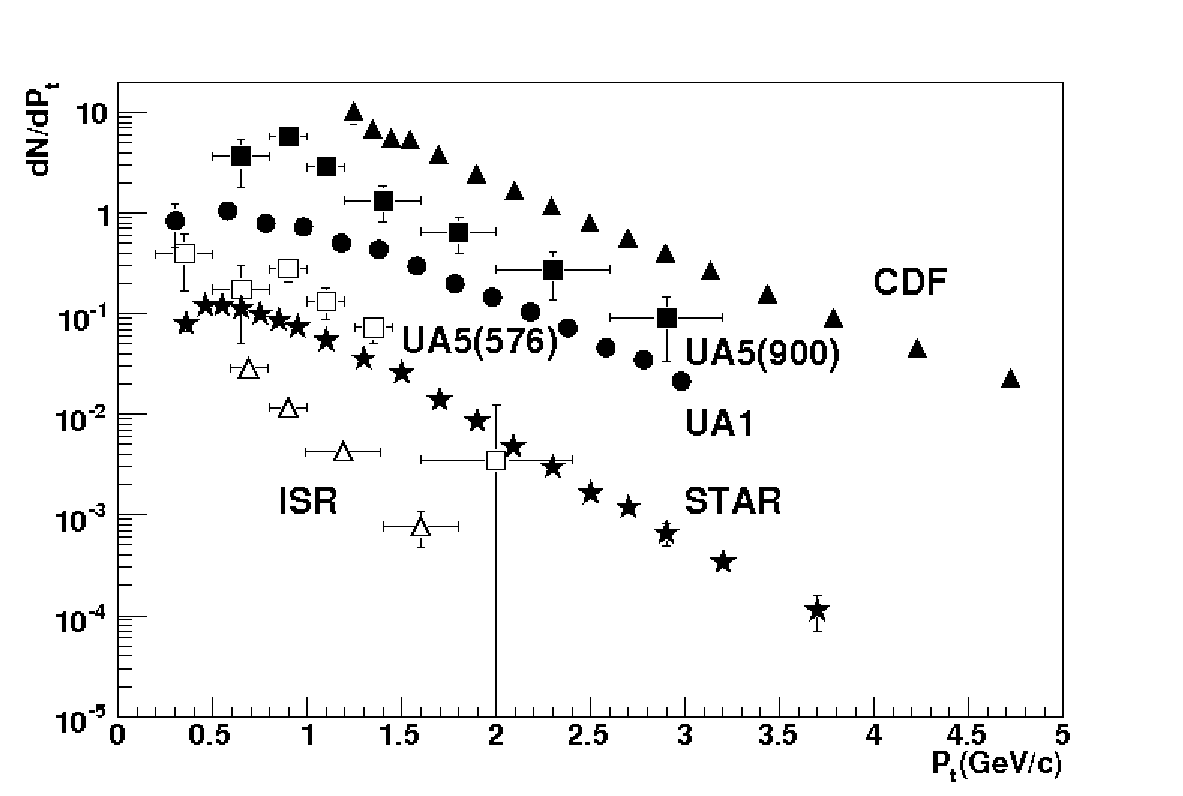}
  \caption{ Transverse momentum distributions of $\Lambda^0$ hyperons from colliders that preceded LHC. The data are from ISR \protect\cite{isr} $p-p$ at $\sqrt{s}= 53 GeV$ - empty triangles, STAR \protect\cite{star} $p-p$ at $\sqrt{s}$=200 GeV - black stars; UA5 \protect\cite{ua5}$\bar{p}-p$ energies : 546 GeV(empty squares) and 900 GeV(black squares); UA1 \protect\cite{ua1} $\bar{p}-p$ (630 GeV) - black circles and CDF \protect\cite{cdf} $\bar{p}-p$ at 1.8 TeV - black triangles.}
  \label{spectracompiled}
\end{figure}

The figure~\ref{spectracompiled} presents the compilation of the data, $dN^(\Lambda^0)/dp_t$, in the region 0.1 GeV/c $< p_t <$ 5 GeV/c from the following experiments: ISR \cite{isr}, STAR \cite{star}, UA1 \cite{ua1}, UA5 \cite{ua5} and CDF \cite{cdf}. It illustrates the changes of hyperon transverse distributions on the energy range from ISR to Tevatron experiments. We do not need the absolute values of distributions in our calculations, hence those were chosen arbitrarily. The range of low $p_t$ 0.3 GeV/c $< p_t <$ 4 GeV/c has the most impact on the value of average $p_t$. The figure clearly shows that average transverse momenta grow with energy.

\section{Preliminary Comparison of Hyperon Transverse Momentum Spectra from LHC Experiments}

The recent data on $\Lambda^0$ hyperon distributions are obtained in the following LHC groups: ALICE \cite{alice} at 900 GeV, ATLAS \cite{atlas} and CMS \cite{cms} at 900 GeV and 7 TeV. 
We are going to compare the results of these LHC experiments with the data of lower energy $p-p$ colliders, ISR \cite{isr}($\sqrt{s}= 53 GeV$) and STAR \cite{star}($\sqrt{s}= 200 GeV$).
  
\begin{figure}[htpb]
  \centering
  \includegraphics[width=12.0cm, angle=0]{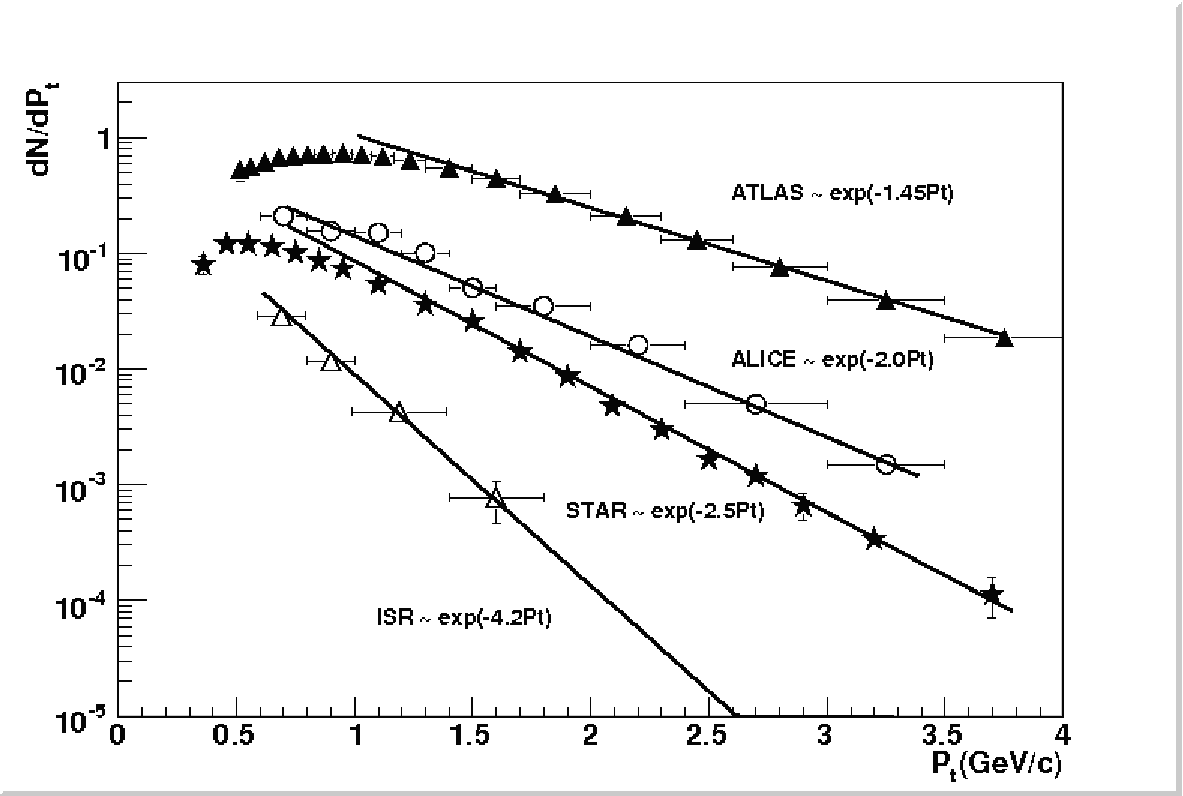}
  \caption{ Transverse momentum distributions from $p-p$ collider experiments: ISR\protect\cite{isr} (53 GeV) - empty triangles, STAR\protect\cite{star} (200 GeV) - black stars, ALICE\protect\cite{alice} (900GeV) - empty circles and ATLAS\protect\cite{atlas} (7 TeV) - black triangles, as fitted with the exponents.}
  \label{isr_star_alice_atlas}
\end{figure}

The slopes of spectra, B, on the figure~\ref{isr_star_alice_atlas} , change, if we fit the data with a simple exponential function: exp(-B*$p_t$).
 
\begin{figure}[htpb]
  \centering
  \includegraphics[width=12.0cm, angle=0]{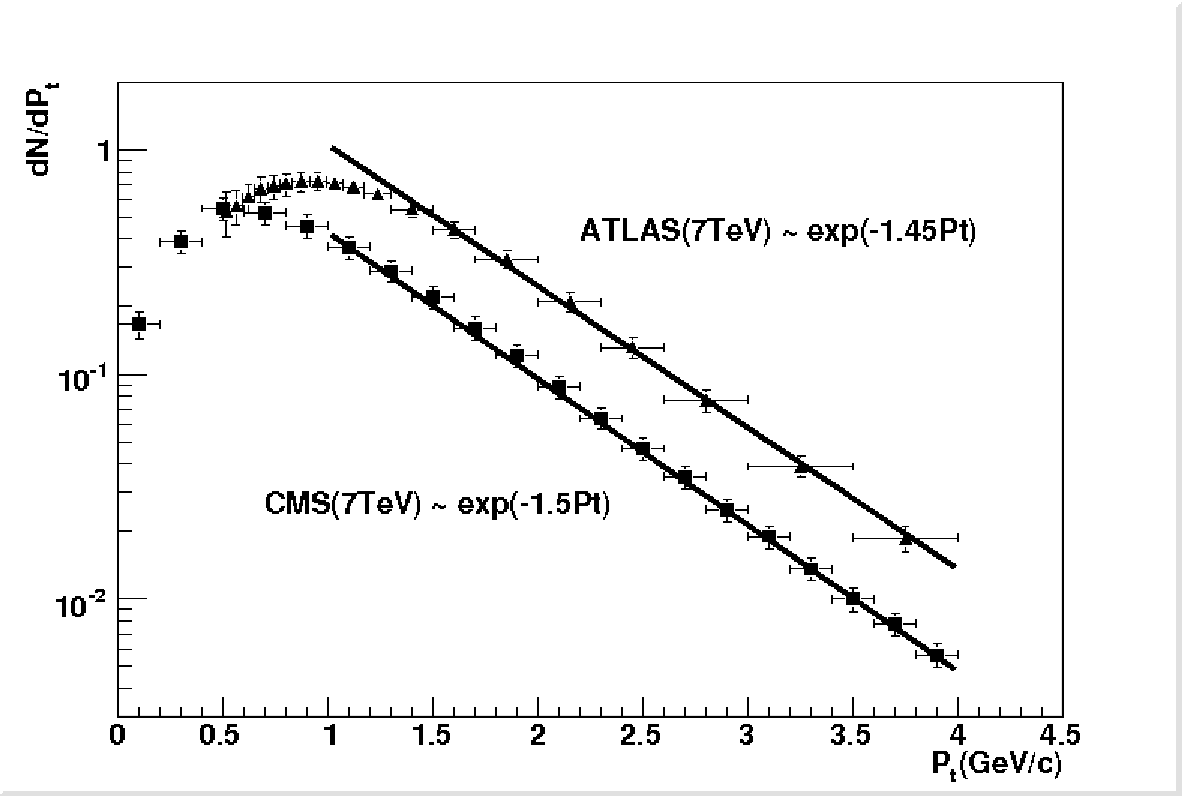}
  \caption{ Transverse momentum distributions at $\sqrt{s}$= 7 TeV: ATLAS \protect\cite{atlas} - black triangles and CMS\protect]cite{cms} - black squares}
  \label{atlas_cms}
\end{figure}

We can conclude that transverse momentum spectra are harder with the energy growth beginning from slope B=4,2 for ISR data, to B=2,6 for STAR and to B=2,0 at 900 GeV for ALICE. The slope is flatter at $\sqrt{s}$ = 7 TeV, B = 1,5.  The calculation of average transverse momenta requires a detailed description of spectra, see section 3.

Both LHC experiments at 7 TeV, ATLAS and CMS, have presented the hyperon spectra with the same slopes as expected  (see the figure~\ref{atlas_cms}). The different forms of the distributions at low $p_t$ region might be caused by efficiency specifics of ATLAS experiment.

\section{Baryon Transverse Momentum Distributions in QGSM}

The QGS Model has been devised for the description of rapidity distributions and hadron spectra in $x_F$ \cite{kaidalov,qgsmtheory,hyperon,heavyquarks,lambdasym,arakelyan}. The Model operates with pomeron diagrams (see the figure~\ref{antipomeron}), which help calculate the rapidity spectra. These spectra are presented as the convolutions of proton constituent quark structure functions with the diquark-antidiquark pair distributions at the pomeron cylinder fragmentation into baryons. This approach took into account mostly the average $p_t$ values for given energy.
The early QGSM study \cite{veselov} on the hadron transverse momentum distributions has shown that the spectra of baryons in proton-proton collisions can be described with the following $p_t$-dependence:

\begin{equation}
E \frac{d^{3}\sigma^H}{dx_F d^{2}p_{t}}= \frac{d\sigma^{H}}{dx_F}*A_0*\exp[-B_0*(m_t-m_0)],\nonumber 
\end{equation}

where $m_0$ is the mass of produced hadron, $m_t$ = $\sqrt{p_t^2+m_0^2}$. The slope parameter, $B_0$, used to bring the dependence on $x_F$, as it was suggested in previous research \cite{veselov}. The values of $B_0$ for the spectra of many types of hadrons ($\pi$, K, p) were estimated for the data of proton-proton collisions up to the energies of ISR experiment. The value of the slopes of baryon spectra for the data in central region of rapidities was universal and equal to $B_0$ = 6,0.  

\begin{figure}[htpb]
  \centering
  \includegraphics[width=12.0cm, angle=0]{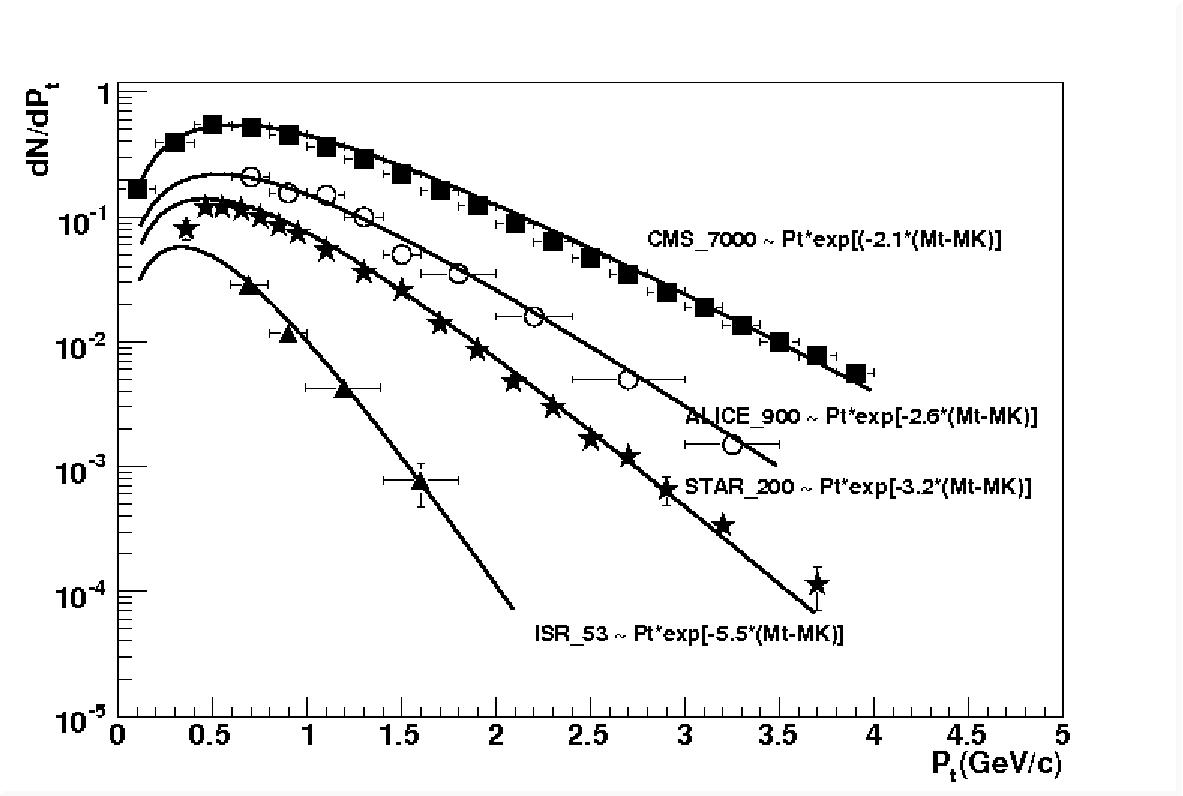}
  \caption{ The description of proton-proton experiment data ISR\protect\cite{isr} (53 GeV), STAR\protect\cite{star} (200 GeV), ALICE \protect\cite{alice}(900GeV) and CMS\protect\cite{cms} (7 TeV) data on hyperon production with the QGSM fit.}
  \label{hyperonspectra}
\end{figure}

As discussed above, the slopes of spectra, $B_0$, at the modern collider experiments depend on energy. Moreover, the form of spectra at LHC and RHIC indicates that the value of $m_0$ is not the mass of proton or hyperon. A better description of hyperon spectra can be achieved with $m_0$ = 0,5 GeV that is actually the mass of kaon, see the figure~\ref{hyperonspectra}. This effect can be provisionally explained as the minimal transverse momentum of hyperon at the fragmentation of diquark-quark chain (see the QGSM pomeron diagram for p-p collisions in the figure~\ref{antipomeron}). The value of $m_0$ should be equal to the kaon mass, because the minimal diquark-quark chain fragmentation produces only two hadrons: $\Lambda^0$+K.

\section{The Difference between Distributions in Proton-Proton and Antiproton-Proton Collisions}

Here we consider the difference in the spectra of baryon production in symmetric ($p-p$) and asymmetric ($\bar{p}-p$) reactions.

\subsection{The Data of UA5 Experiment}

\begin{figure}[htpb]
  \centering
  \includegraphics[width=10.0cm, angle=0]{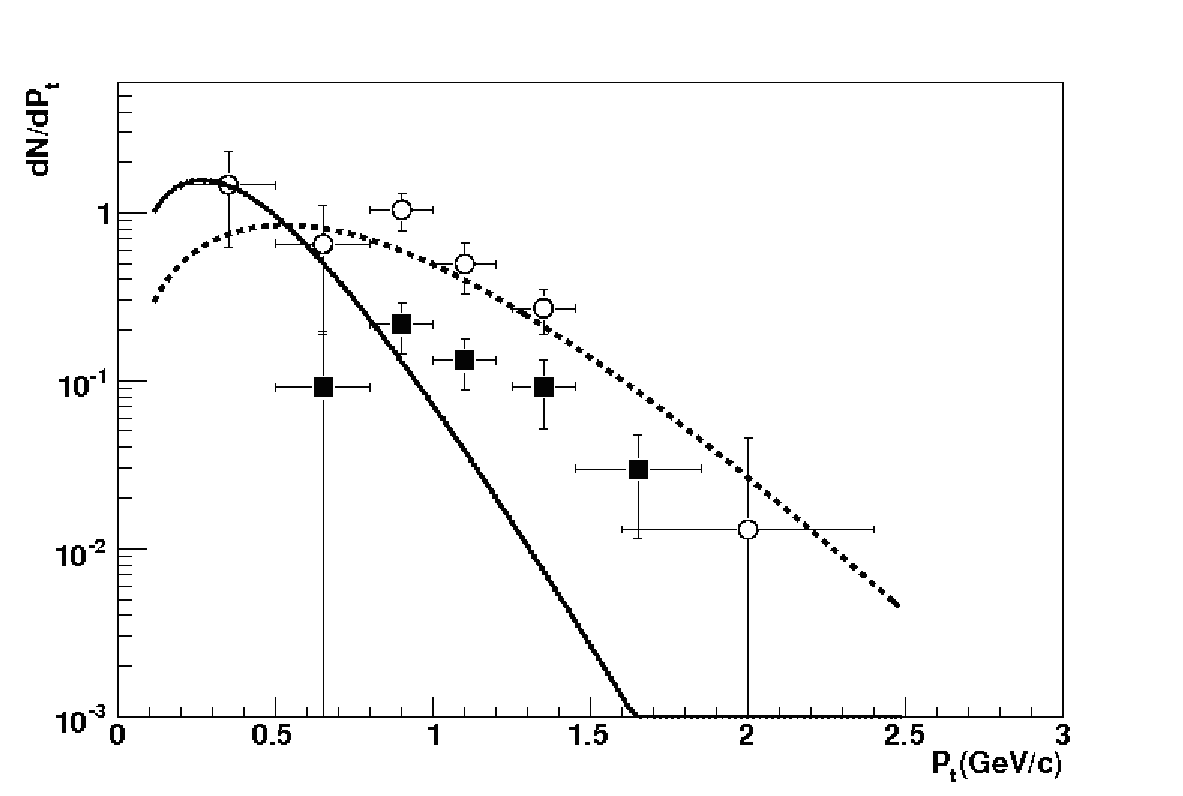}
  \caption{The forms of spectra at antiproton-proton reaction in UA5\protect\cite{ua5} - black squares at the  energy $\sqrt{s}$=200 GeV. The UA5\protect\cite{ua5} data of higher energy ($\sqrt{s}$=546 GeV) are shown with empty squares.  As I mentioned above, the absolute values of spectra are chosen arbitrarily. The fit for UA5(546) with solid and dashed lines demonstrates two different components at the asymmetric one-pomeron cut.}
  \label{ua5fit}
\end{figure}

This subsection discusses the influence of quark composition of beam particles on the shape of transverse momentum spectra of $\Lambda^0$ hyperon production. The data from $p$-$\bar{p}$ experiments UA5 \cite{ua5} of energies, $\sqrt{s}$= 200 GeV and 546 GeV, are studied (see figure~\ref{ua5fit}).

The sharp exponential contribution to spectra is seen in $\bar{p}-p$ reaction at $\sqrt{s}$ = 546 GeV in UA5 collaboration data\cite{ua5} at very low $p_t <$ 0.5 GeV/c, see the figure~\ref{ua5fit}.  This exponential component might exist in other antiproton-reaction spectra as well, but it is not seen because of the absence of measurements in low $p_t$'s.

The form of the spectrum at low $p_t$ has a strong impact on the value of cross section, if the experimental distributions are integrated beginning from low momenta, $p_t$ = 0.3 GeV/c.
The resulting cross section from antiproton-proton reaction should be smaller than the cross section, obtained in proto-proton collision of the same energy, if there are no data points at $p_t <$ 0.5 GeV/c. The complex form of distribution in UA5 can be explained by two components in the spectra for antiproton-proton collisions. The QGSM diagrams explain the nature of two components in the asymmetrical reaction.
 
\subsection{The Difference in Distributions from antiproton-proton and proton-proton reactions in QGSM}

The difference in $p_t$-spectra of $\Lambda^0$'s produced in high energy $p-p$ and $\bar{p}-p$ collisions cannot be explained in the perturbative QCD models. On the other hand, total cross sections  in $p-p$ and $\bar{p}-p$ collisions should be equal because they depend only on the parameters of the Pomeron exchange between two interacting hadrons and should not be sensitive to the quark contents of colliding particles. It seems, the difference exists only in the form of spectra.

\begin{figure}[htpb]
  \centering
  \includegraphics[width=12.0cm, angle=0]{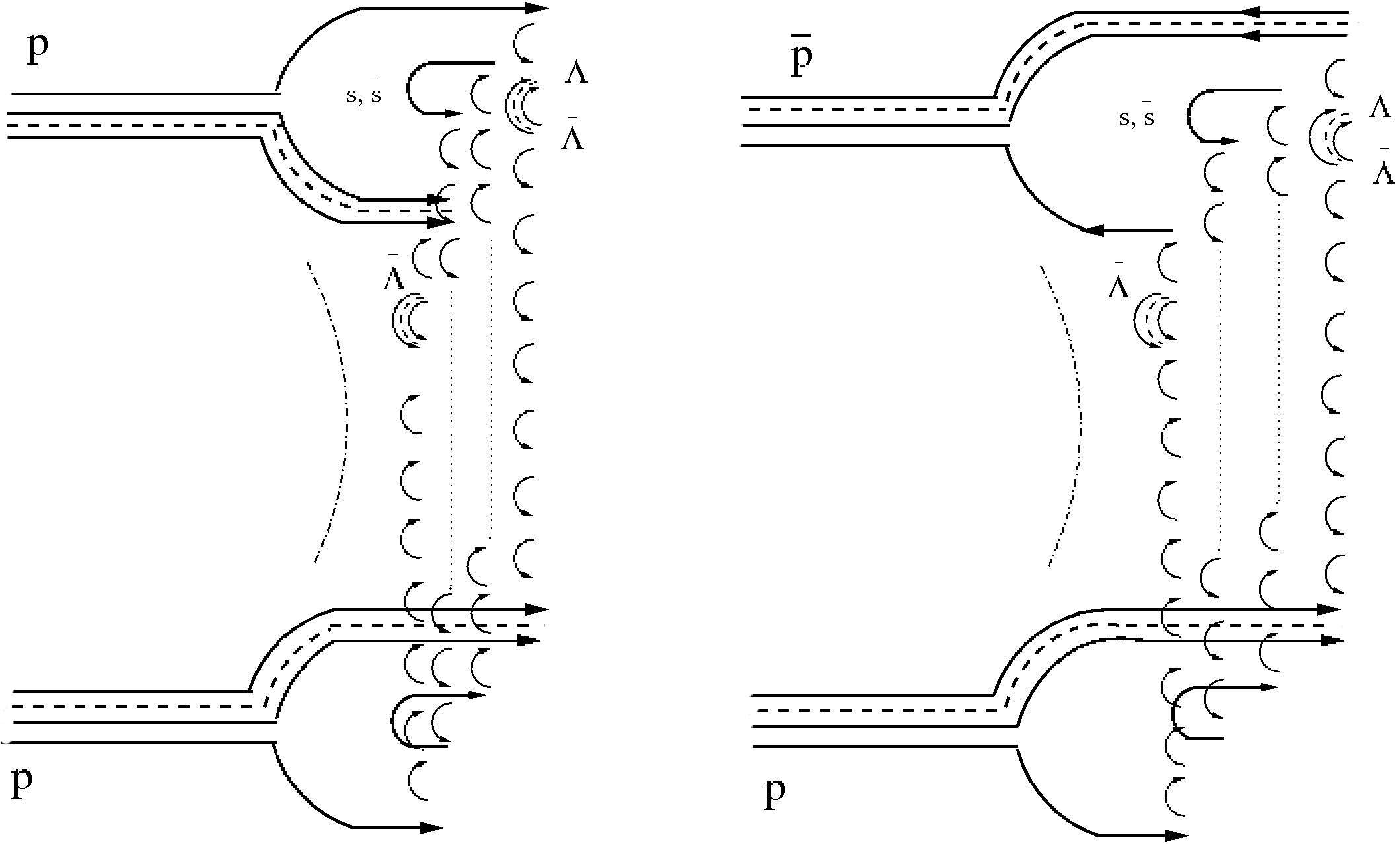}
  \caption{QGSM multiparticle production diagrams for a) proton-proton and b) antiproton-proton reactions.}
\label{antipomeron}	
\end{figure}

\begin{figure}[htpb]
  \centering
  \includegraphics[width=8.0cm, angle=0]{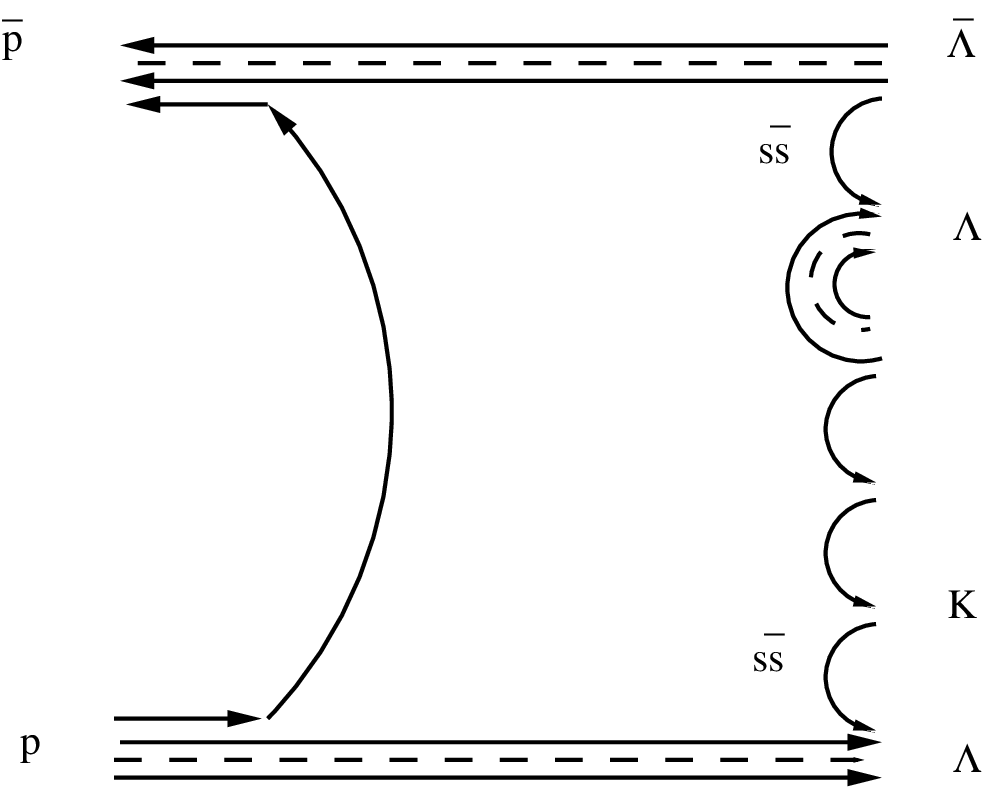}
  \caption{The low energy diagram with quark-antiquark annihilation and diquark-antidiquark chain fragmentation.}
\label{lowenergy}	
\end{figure}

The pomeron diagrams of $p-p$ and $\bar{p}-p$ collisions are shown in the figure~\ref{antipomeron}.
In the framework of QGSM \cite{qgsmexp,hyperon,veselov,heavyquarks,protonantiprotondiff}, which is based on the Regge theory and on the phenomenology of pomeron exchange, the spectra of produced particles at low energies are the results of the cut of one-pomeron exchange diagram. The comparison of diagrams shows that the most important contribution to hadron production spectra in $\bar{p}-p$  reaction is brought with the fragmentation of antidiquark-diquark chain of pomeron cylinder, because this side of diagram certainly takes the greater part of energy of colliding particles. Otherwise, the $p-p$ collision diagram is symmetric and built from two similar quark-diquark chains. It seems that just the incorrect description of the region of small transverse momenta at antiproton-proton reaction leads to underestimated values of cross section for the low energy collider experiments. With the growing energy, the spesifics of asymmetric one-pomeron diagram will not be seen because of dominating contributions of multi-pomeron exhanges that are symmetrical \cite{lambdasym,averagept}.
The figure~\ref{lowenergy} shows that baryon production at low energy goes with the quark-antiquark annihilation. The resulting spectrum consists of  the contribution from only diquark-antidiquark chain that allows us to see the pure form of baryon transverse momentum distributions in asymmetric case of fragmentation. The form of baryon spectra in antiproton-proton reaction of low energy has been planned to be studied in experiment TAPAS \cite{tapas}. 

Since LHC experiments have proton projectiles, the experimental transverse momentum distributions of hyperons should seem "softer" in the comparison to the predictions of MC generators \cite{heralhc,montecarlo,cosmicray}, which have been preliminary tuned to Tevatron data. 

\section{Average Baryon Transverse Momenta vs. Energy and "Knee" in Cosmic Ray Spectra}

As the data on baryon distributions in antiproton-proton reactions are irrelevant for consideration, the spectra from LHC can be compared only with measurements of proton-proton collision experiments : ISR and STAR. We consider the transverse momentum spectra in the wide range of energies: beginning from 53 GeV to the LHC energy 7 TeV.

\begin{figure}[htpb]
  \centering
  \includegraphics[width=12.0cm, angle=0]{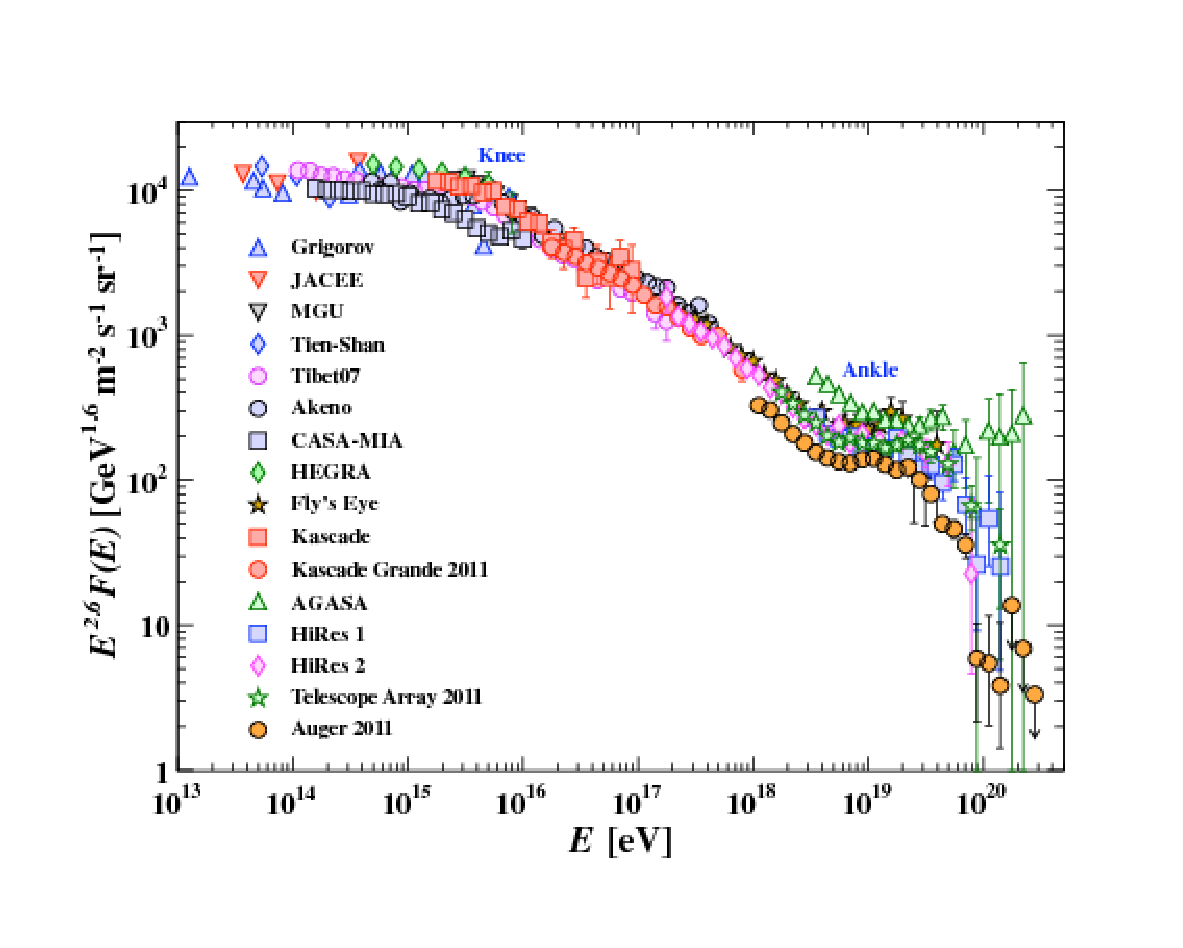}
  \caption{The energy spectrum of cosmic particles that are mostly protons is the result of data compilation from many experiments. The peculiarities in spectrum as "knee" and " ancle" are worldwide-accepted features, where "knee" corresponds to the energy of protons in c.m.s. energy range of LHC  \protect\cite{knee}. Experimental distributions are multiplied by $E^{2.6}$ for convenience, while the original CR spectrum has the slope -2.6.}
  \label{knee}
\end{figure}

The cosmic ray proton spectrum shows the "knee" (see figure~\ref{knee}) at the energy gap between Tevatron and LHC colliders \cite {knee}. 
The change in the slope of proton spectrum at $E_{lab}$ $\approx$ 3* $10^{15}$ eV might be a manifestation of a new regime in hadronic interactions.  Otherwise, the change in the spectrum  has an astrophysical genesis. The cosmic protons, which energy extends far than $E=10^{20}$ eV, may be produced in relativistic jets from Super Massive Black Holes \cite{torus}.
 
\begin{figure}[htpb]
  \centering
  \includegraphics[width=12.0cm, angle=0]{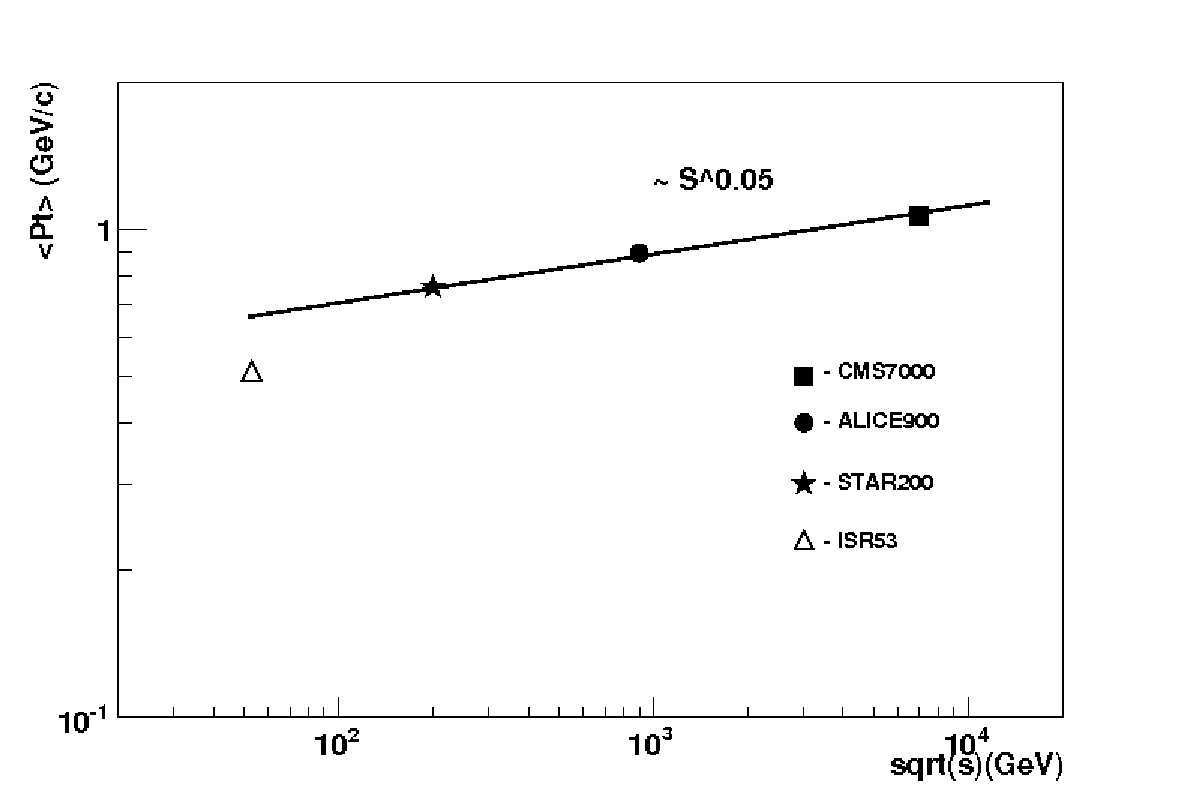}
  \caption{ Average transverse momenta of hyperons increase with the energy, as $s^{0.05}$.}
\label{averageVSenergy}	
\end{figure}

The analysis of baryon transverse momentum distributions in the framework of QGSM \cite{heralhc,lambdaknee,cosmicray} clearly demonstrates that the average transverse momentum of baryons grows steadily with energy (see figure~\ref{averageVSenergy}). 

The $<p_t>$ values grow in the range of energy from 53 GeV (ISR) up to $\sqrt{s}$ = 200 GeV (STAR) and then they go with the asymptote $s^{0.05}$ \cite{bylinkin,hadronmass}. This behavior cannot be considered as substantial change in hadroproduction processes and the certain energy. Since no spesific points exist in  baryon production up to $\sqrt{s}$ = 7 TeV, which corresponds to $E_{lab}= 2,5*10^{16}$ eV, it is reasonable to conclude that "knee" has an astrophysical explanation. 

The growth of average transverse momenta was calculated in the framework of  QGSM on the energy distance up to LHC. The expected QGSM explanation is the following: the dependence on energy has been seen in the average transverse distribution, because in our model the slopes of transverse momentum spectra still brings a little dependence on $x_F$ of the multi pomeron chains that can be short in N-pomeron exchanges . This result should be used for the interpretation of futher data of LHC groups. However, the energy dependence of $<p_t>$ is not yet included into MC generators calculations. In such a way, the results of this research will help improve LUND, Pythia and other MC models.

\section{Average Transverse Momenta of Heavy Hadron Production at LHC.}

\begin{figure}[htpb]
  \centering
  \includegraphics[width=12.0cm, angle=0]{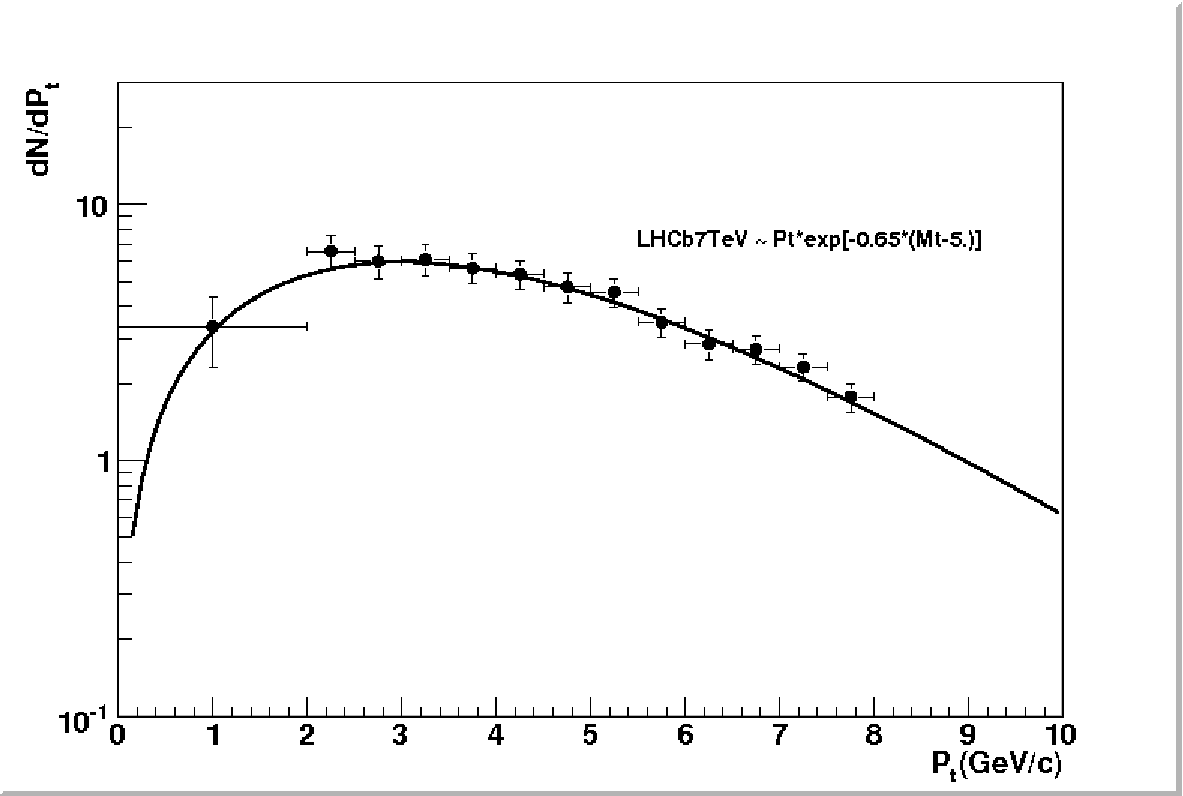}
  \caption{The QGSM fit of B-meson spectra at $\sqrt{s}$=7 TeV in LHCb\protect\cite{lhcbmeson} experiment.}
\label{HQspectra}	
\end{figure}

The previously published analysis of transverse momentum spectra of baryons from LHC experiments (ALICE, ATLAS, CMS) \cite{bylinkin} provided only partial data on hadron spectra. In order to get a full understanding of average transverse dependence at heavy hadron masses, we supplement here the data with the spectra of  D-mesons and B-mesons from LHC \cite{lhcbmeson} at 7 TeV. The heavy quark meson spectra were fitted with the same formula (1) as the baryon spectra (see the figure~\ref{HQspectra}).

After this fit, the average $p_t$ of B-mesons have been estimated as 4,46 GeV at the energy $\sqrt(s)$=7 TeV that is almost equal to the mass of heavy B-hadrons.
The mass dependence of average transverse momenta of hadrons are discussed in my recent papers \cite{hadronmass,recent}.

\section{Conclusions}

The overview of results in transverse momentum distributions of hyperons produced in proton-proton collisions of various energies \cite{lambdaknee,hadronmass} has revealed a significant change in the slopes of baryon spectra in the region of $p_t$ = 0,3 - 8 GeV/c. The spectra of baryons become harder with the energy growth from ISR($\sqrt{s}$=53 GeV) and RHIC (200 GeV) up to LHC (0,9 and 7 TeV). The detailed analysis of hyperon spectra in the framework of QGSM demonstrates the change of slopes from $B_0$ = 4,6 (ISR at 53 GeV) to $B_0$ = 2,1 (LHC at 7 TeV). 
The transverse momentum baryon spectra in antiproton-proton collisions (UA1, UA5, CDF) differ from the $p_t$ distributions of baryons in proton-proton collisions (ISR, STAR, LHC).

QGSM explains this phenomenon as a difference in the splitting of transverse energy between two sides of pomeron diagram. The cut diagram for antiproton-proton case includes the unusual side with the diquark-antidiquark ends, which accumulates more energy than another quark-antiquark side of cylinder. It is reasonable to suggest that the difference in spectra disappears with the growth of energy due to the increasing multi-pomeron contributions into the differential cross section that are similar in this case for both  $\bar{p}-p$ and $p-p$ collisions.

The average $p_t$ values in proton-proton collisions grow steadily as the power of energy, $s^{0.05}$, up to highest LHC energy 7 TeV.  
We did not see the dramatic change in the baryon production processes at the energies of LHC. This conclusion has crucial implications for cosmic ray physics, since it suggests that the "knee" at $E_{lab}$ $\approx$ 4* $10^{15}$ eV in cosmic proton spectra does not originate in hadronic interactions. There is no significant change in baryon spectra up to $\sqrt{s}$ = 7 TeV corresponding to $E_{lab}= 2,5*10^{16}$ eV in the cosmic proton spectrum. 

The QGSM explanation for the growing of average transverse momenta  is the following: the dependence on energy has been seen in the average $p_t$, because in our model the slopes of transverse momentum spectra still bring a little dependence on $x_F$ and multi pomeron chains  can be short in multi-pomeron exchanges due to the energy conservation.

The prediction of transverse momentum behavior is useful for the planning of  future colliders of higher energies.
The idea of proton production in space warrants a further detailed  investigation of the hadroproduction dynamics in the framework of QGSM.

\section{Acknowledgments}
 
I would like to thank the Russian Foundation of Fundamental Research (grant 13-02-06091) and Dmitry Zimin's "Dynasty" Foundation. Their financial supports made possible the First Kaidalov's Phenomenology Workshop. This workshop has given the boost to the phenomenological research that is described in my paper.

\end{document}